\documentclass[sigconf]{acmart}
\usepackage{graphicx}
\usepackage{hyperref}
\usepackage{xcolor}
\usepackage{amsmath}
\usepackage[export]{adjustbox}

\graphicspath{{figs/}}

\begin{document}
  \title{Beer2Vec : Extracting Flavors from Reviews for Thirst-Quenching Recommandations}

\author{Jean-Thomas Baillargeon}

\email{jean-thomas.baillargeon@ift.ulaval.ca}

\affiliation{%
  \institution{Université Laval}
  \city{Québec}
  \country{Canada}
}

\author{Nicolas Garneau}

\email{nicolas.garneau@ift.ulaval.ca}

\affiliation{%
  \institution{Université Laval}
  \city{Québec}
  \country{Canada}
}

\renewcommand{\shortauthors}{Baillargeon et al.}

\ccsdesc[500]{Information systems~Retrieval models and ranking}
\ccsdesc[500]{Computing methodologies~Artificial intelligence}

\begin{abstract}
This paper introduces the \textit{Beer2Vec} model that allows the most popular alcoholic beverage in the world to be encoded into vectors enabling flavorful recommendations.
We present our algorithm using a unique dataset focused on the analysis of craft beers. 
We thoroughly explain how we encode the flavors and how useful, from an empirical point of view, the beer vectors are to generate meaningful recommendations.
We also present three different ways to use \textit{Beer2Vec} in a real-world environment to enlighten the pool of craft beer consumers.
Finally, we make our model and functionalities available to everybody through a web application. 

\keywords{Representation Learning \and Recommendation \and beer}
\end{abstract}

\setcopyright{acmcopyright}
\copyrightyear{2022}
\acmYear{2022}
\acmDOI{XXXXXXX.XXXXXXX}

\acmConference[NLPIR '22]{}{December 16--18,
  2022}{Bangkok, Thailand}
\acmISBN{978-1-4503-XXXX-X/18/06}
\acmPrice{15.00}

\maketitle

\section{Introduction}
A decade ago, selecting a beer from the liquor store was an easy task. Major companies were offering a limited array of products from large scale production. Nowadays, an influx of craft beers offers new tasting experiences to amateurs, and many brewers are {\em tapping} into this market. Ten years ago, there were 310 Canadian breweries\footnote{\url{https://industry.beercanada.com/statistics}} competing for the average Canadian palette. Today, over 1100 breweries\footnote{\url{https://en.wikipedia.org/wiki/Beer_in_Canada}} compete, not only for the average amateur money but also for rare product niches, whose price range overlaps with the ones of good wine bottles.

This vast offer isn't without drawbacks for both sides of the value chain. On the one hand, it overwhelms the consumer with too many hard-to-distinguish products. On the other hand, the sellers may not have a beer connoisseur readily available to help the customer find the most appropriate beer among the ones he has in-store.

Rating systems such as {\bf untappd\footnote{\url{https://untappd.com/}}} or {\bf ratebeer\footnote{\url{https://www.ratebeer.com/}}} provided important tools for beer geeks around the world to distinguish a beer (or a brewery) from its peer. These systems allow them to assess the quality--to some extent--of the next beer they will drink and enable them to share their beverage appreciation. This exchange of information is made on the promise of an enhanced recommendation for their next beer. 

These recommendation systems, however, suffer from three main problems. They use a similarity ranking based on user preference proximity that forgoes the beer's flavor profile aspect. These systems don't consider the purchase context, where a user needs to select a beer from a limited amount of products or how a user would like to select a beer based on a personal choice of flavors.
Moreover, some clients may have difficulty expressing the flavors they are looking for.

We provide a solution for these three problems by introducing the {\em Beer2vec} model, a vector representation learning procedure that pulls its roots from the word2vec's skip-gram model\cite{mikolov2013efficient}.
{\em Beer2vec} naturally enables vector arithmetics, allowing beers to be expressed in terms of flavors. Transforming beers into vectors implies that a user can select an excellent beer to his liking when faced with an immense wall of beer by comparing two beers' taste profiles with a straightforward vector distance calculation. He can obtain the optimal beer from the shelf without requiring the merchant to hire a beer sommelier. Furthermore, the vectorial operation allows analogies in the {\em Beer2vec} application, where a user can add and subtract tastes from a beer to interpolate towards another one. The code and the beer vectors will be publicly available upon acceptance.

The {\em Beer2vec} application is also showcased on a website that contains all the experimentations and results from Section~\ref{sec:exps}. Although the scope of this work is limited to beer, we believe that other communities using flavor tagging, such as \textbf{Vivino}\footnote{\url{https://www.vivino.com/}} for vine amateurs and \textbf{Connosr}\footnote{\url{https://www.connosr.com/}} for whisky amateurs would benefit from this application.

The next sections are organized in the following way; we first introduce in Section~\ref{sec:related_work} the related work regarding representation learning for recommendations, with a specific scope in the nutrition field.
We then present in Section~\ref{sec:dataset_creation} how we created the dataset needed to learn flavorful vector representations, followed by the training methodology and architecture in Section~\ref{sec:model}.
In Section~\ref{sec:exps}, we further describe the experiments we conducted to assess our vectors' palatability.
We conclude this paper with possible future works in Section~\ref{sec:conclusion}.

\section{Related Work}
\label{sec:related_work}

Recent advances in learning dense representations for words are using shallow
neural networks such as GloVe\cite{Pennington2014GloveGV} and word2vec\cite{mikolov2013efficient}.
The latter model's name and architecture led to a plethora of *2vec models~\footnote{As listed here; \url{https://gist.github.com/nzw0301/333afc00bd508501268fa7bf40cafe4e}, there are 130 and counting *2vec models.}.
From these models, we can find vector representations for code, emojis, time, food, and even names\cite{Foxcroft2019Name2VecPN}, to name a few.

It has been shown that vector representation learned by following a training scheme similar to what \cite{mikolov2013efficient} proposed conveys a certain analogical reasoning~\cite{Mikolov2013DistributedRO} i.e., the possibility to perform vector arithmetics (to some extent~\cite{Levy2014regularities, Linzen:16, Rogers:ea:17}).
Nonetheless, vector representations are deemed useful in recommender systems by capturing semantic similarity among terms in a vocabulary~\cite{Rudolph2016ExponentialFE}.
The learning procedure has also been applied to learn metadata embeddings to provide better cold-start recommendations for users amongst a set of items~\cite{Kula2015MetadataEF, Gspr2019ImprovingTP}. We believe this recommendation technique can improve the idea of food recommendations presented in \cite{el2012food}.

However, while being probably the alcoholic beverage the most consumed across the world, we could not find vector representations dedicated to the beer universe.
This absence is perhaps due to the lack of datasets for this specific case.
We thus introduce in the following section how we created such as dataset to train yet another flavor\footnote{Pun intended.} of word2vec, \textit{Beer2vec}.






\section{Dataset Creation}
\label{sec:dataset_creation}

The dataset we used in the experimentation was obtained from {\bf untappd} in 2018 and consists of check-ins, the user's appreciation of a beer. The excerpt of data contains a total of 500 000 check-ins for over 14600 distinct beers.


The check-ins represent user input for a specific beer. The check-in contains metadata to identify the beer, such as the beer name and the brewer's name. The check-ins also include information regarding the user's beer appreciation, such as a text review and a tasting score, but more importantly, a list of standardized flavor tags drawn from a list of 264 predetermined elements. These tags indicate the positive and negative aspects of the many facets of beer taste, namely malt, hop, and yeast. In our experiment, we only used the flavor tags since they are rich and easy to use.

We evaluated the feasibility of using the text portion of the review. However, after inspecting the quality of the written reviews, we had an important doubt and decided not to consider them in our analysis.

\section{\textit{Beer2Vec}}
\label{sec:model}

\textit{Beer2Vec} is heavily inspired by the Skip-Gram architecture introduced by \cite{Mikolov2013DistributedRO}.
Given the set of beers $\mathcal{B}$ and the set of flavors $\mathcal{F}$,
\textit{Beer2Vec} is parameterized by two embedding matrices;
$\Theta = \mathbf{B} \in \mathbb{R}^{|\mathcal{B}| \times k}, \mathbf{F} \in
\mathbb{R}^{|\mathcal{F}| \times k}$, which are the beers and flavors
embeddings respectively, and where $k$ is a hyperparameter setting the dimension of both matrices.

Given the set of check-ins $\mathcal{C}$, that constitutes our training set, where every check-in $c_i$ contains a beer $b_i$ and a set of flavors $f_i$, the \textit{Beer2Vec} model tries to maximize is the average log probability of the following function;

\vspace{0.2cm}
\begin{equation}
  \mathcal{J} = \frac{1}{|\mathcal{C}|} \sum_{i=1}^{|\mathcal{C}|} \sum_{j=1}^{|f_i|} \text{log } p(f_{i,j}|b_i).
\end{equation}
  \vspace{0.2cm}

More concretely, \textit{Beer2Vec} learns to predict flavors $f_{i,j}$ given a beer $b_i$.
Similar to \cite{Mikolov2013DistributedRO}, $p(f_{i,j}|b_i)$ is expressed using the softmax function and we thus optimize the following loss function w.r.t to the parameters $\Theta$;

\vspace{0.2cm}
\begin{equation}
  \mathcal{J}(\Theta) = \frac{1}{|\mathcal{C}|} \sum_{i=1}^{|\mathcal{C}|} \sum_{j=1}^{|f_i|} \frac{\text{exp}\left( \mathbf{b}_i ^{\intercal} \mathbf{f}_{i,j}\right)}{\sum_{m=1}^{|\mathcal{F}|} \text{exp}\left( \mathbf{b}_i ^{\intercal} \mathbf{f}_{m}\right)}.
\end{equation}
  \vspace{0.2cm}

In practice, we use a dimension $m$ of $5$, Adam as the optimizer with a standard learning rate of $0.001$, a batch size of $128$, and train the model for a maximum of $300$ epochs.

\section{Applications}
We present three applications of \textit{Beer2Vec} in terms of recommendation and information retrieval. The first one allows customers to select the most appropriate beverage while visiting a beer store. The second allows him to search for beer based on a custom-made flavor profile. The last one allows a person drinking a beer to obtain the correct words to describe its beer.

\subsection{Select a Beer when Visiting a beer store}
Whenever this customer visits a beer outlet and requires a recommendation, he simply compares the vector representations (using the dot product) of available beers and the ones from his favorite beers. The customer sorts the available beers using the calculated distance metric to obtain the most similar beverages compared to his favorite list, hence providing a \textit{flavorful recommendation}.
For example, let $\mathbf{B}_i$ be the current set of beers available in the outlet and $\mathbf{U}_j$ be the customer's set of favorite beers, a customer can obtain the $n$ most interesting beers using the following retrieval formula;
\begin{equation}
    b_j = \underset{n}{\text{arg\_sort}}(\textbf{B}_i \times \textbf{U}_j^\top)
\end{equation}

\subsection{Search Beer by Flavor Profile}
The second way of obtaining a recommendation is by creating a flavor vector based on its own preferences. In this application, a user selects flavors, hereby identified as $\mathbf{f}$, and uses vector arithmetics to combine them. One easy way to combine such vectors is by averaging \textit{Beer2Vec} vectors using custom weights to create something unique. Finally, using the dot product, the user compares its created flavor profile vector with all the beers available in the beer2vec database and sorts by distance to find the one with taste the more similar to its request. Similar to the previous use case, we obtain the set of desired beers using the following retrieval formula;
\begin{equation}
    b_j = \underset{n}{\text{arg\_sort}}(\textbf{B}_i \times  (\sum_{i=1}^{|\textbf{f}|} \alpha_i \cdot \textbf{f}_i^{\top})),
\end{equation}


with $\sum_{i=1}\alpha_i=1$.

\subsection{Obtain Vocabulary to Describe a Beer}
Another way to exploit the \textit{Beer2Vec} model is to use it to decompose a beer vector into its flavor components. This allows a user to compare the vector representation of the beer with the flavors vector in a database using the dot product. Ranking the flavors by their proximity enables the user to express its beverage with connoisseur vocabulary.
Working the other way around, we retrieve the set of $n$ flavors given a beer's vector representation $\mathbf{b}_j$;
\begin{equation}
    f_i = \underset{n}{\text{arg\_sort}}(\mathbf{F}_i \times \textbf{b}_j^{\top})
\end{equation}

\section{Model Evaluation, Results and Analysis}
\label{sec:exps}

We evaluate our model on two axes. The first evaluation axis regards the comparison to the principal component analysis (PCA), our baseline algorithm. The second axis of evaluation is the relevance of the vectors in recommendation quality, amount of flavor embedded in the beer embedding, and how these embeddings are robust to mathematical operations. Taste is very subjective by nature and a seasoned homebrewer with over ten years of serious experience will qualitatively evaluate the recommendations' quality.

\subsection{Taste Vector Visualisation}

\begin{figure}[h!!!!!!!]
    \centering
    \includegraphics[width=0.47\textwidth]{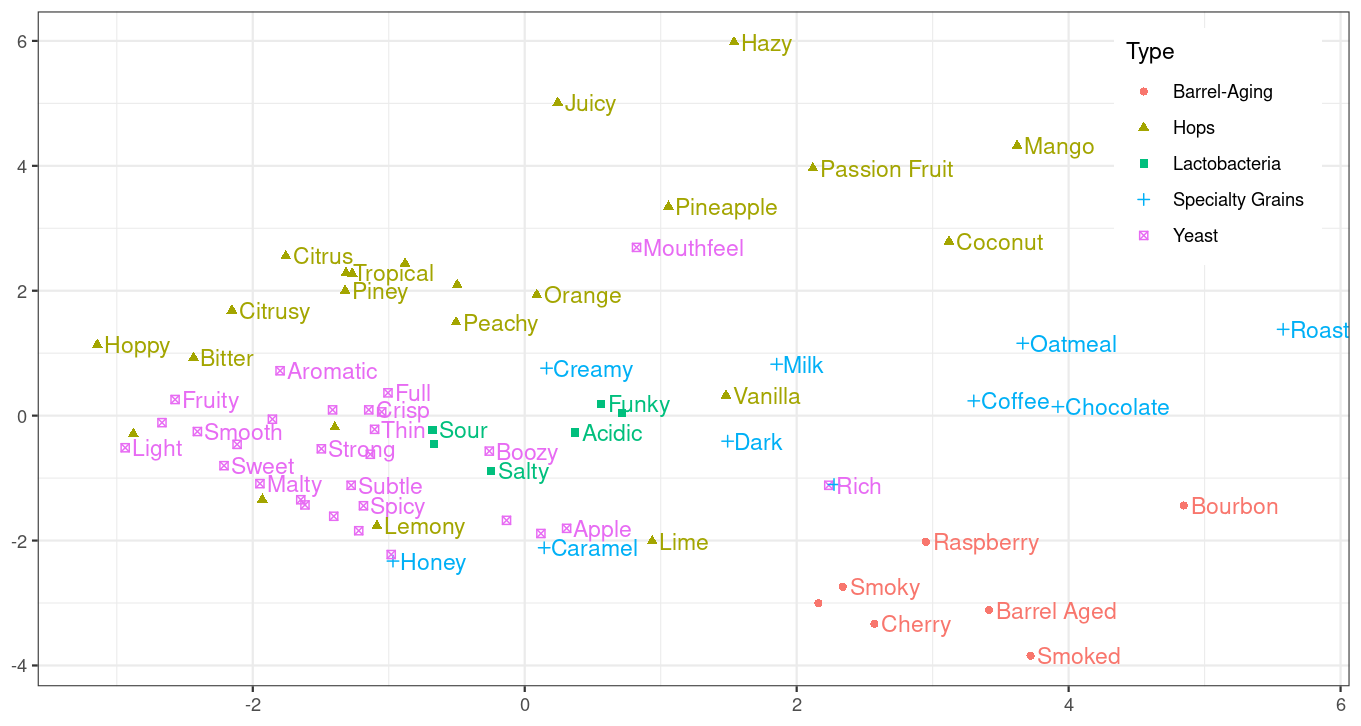}
    \caption{A 2D Projection of the learned flavor matrix over our training set, $\mathbf{F}$.
    We can identify several clusters of flavors that interpolate from one to another.}
    \label{fig:2d_taste_vector}
\end{figure}

To give us a qualitative sense of the embeddings learned, we projected the $\mathbf{F}$ matrix in a two-dimensional plot in Figure~\ref{fig:2d_taste_vector} using the principal component analysis. The vector projection presented generates a flavor map that represents the flavor in beer in a (surprisingly) realistic way. We clustered the taste by their source using marker shapes. The analysis presented below presents how our taste vectors capture the information.

The first element to observe is the large flavor spectrum brought by hops in beer. The transition is shown along the x-axis: on the left side are sharper aspect of the hop notes (\textbf{citrus}, \textbf{grapefruit}, \textbf{orange}) and on the right side are  the rounder flavors (\textbf{pinnaple}, \textbf{passion fruit}, \textbf{mango}, \textbf{coconut}). This wide array of flavors corroborates the idea that hops bring the most taste to the beer.

The second element one notices is how the yeast impact is condensed. This observation reflects another expected taste in beer: yeast flavor, except for sours and Belgian beer, should be tamed. Otherwise, those taste would be seen as off-flavors (\textbf{phenolic}, \textbf{medecine}, \textbf{apple}, \textbf{boozy}), and are outside the yeast impact region. 

The third element to observe is the barrel-aging effect that is isolated and separated from the other flavors. This is not surprising as barrel-aging is not a common practice in beer making and is an additional process that any beer can go through. However, beers that are the most improved by barrel-aging are \textit{stouts}. \textit{Stouts} use a lot of specialty grain that imparts flavors such as \textbf{Oatmeal}, \textbf{roasted}, \textbf{coffee}, \textbf{chocolate}. These flavor types are the closest to the barrel-aging region and enhance our confidence in our methodology.



\subsection{Baseline comparison}
The baseline we selected in our experiment is the PCA, an unsupervised algorithm that extracts the most salient information into vectors. Comparing our method to this baseline is natural since both approaches aim to encode data into dense vectors. We use the same amount of components as our model's embeddings size (i.e., 5) from Section \ref{sec:model}. The individual items for the input matrix $\mathbf{C}$ are given by 
$$ \mathbf{C}[x,y] = \mathrm{count}(f_y \in \mathcal{C}_x), $$
where $f_y$ is the $y^{th}$ flavour in $\mathcal{F}$ and $\mathcal{C}_x$ the subset of check-in that contains the beer $b_x$, the $x^{th}$ beer from $\mathcal{B}$.

To compare against our method, we selected three beers that cover a wide array of flavors, and we found the three most similar beers for both models, as illustrated in Table~\ref{tab:baseline_analysis}.
The selected beers are an IPA ({\bf juicy} and \textbf{hoppy}), a Pilsner (\textbf{crisp} and \textbf{aromatic}), and a stout (\textbf{malty} and \textbf{sweet}).
The similarity metric used in this context is the dot product between $\mathbf{b}_i$ and $\mathbf{b}_j$, the vector representation of beer $b_i$ and $b_j$. The three beers with the highest similarity score are presented for each approach.

\begin{table}
    \centering
    \begin{tabular}{c|c|c}
    \toprule
        \textbf{Beer} & \textbf{PCA} & \textbf{\textit{Beer2Vec}}\\
        \midrule
        Catnip & Cucumber (Sour) & HYPA \#6 (IPA)\\
        (IPA) & Xtra-Citra (IPA) & Dipa Australe (IPA) \\
         & Xingu Black (Lager) &  Cyclope (IPA) \\
         \midrule
         Pilsner Czech & Esperanza (IPA) & Joufflue  (Wheat) \\
         (Pilsner) & Espionne  (Stout) & Jugement (Lager) \\
         & Lorna Bray (Stout) & Custom A (Pilsner) \\
         \midrule
         S.I.R. Popov & Superpause (IPA) & Écossaise (Scotch Ale) \\
         (Stout) & La Kirke (Bitter) &  Tête Brûlée (Stout)\\
         & Krill (Lager) & SpécialB (Belgian Quad) \\
         \bottomrule
    \end{tabular}
    \vspace{0.2cm}
    \caption{Most Similar Beer According to Different Models. Given a Beer, acting as the query, we retrieved the top 3 similar beers with respect to the corresponding model (PCA or \textit{Beer2Vec}).}
    \label{tab:baseline_analysis}
\end{table}

As one can see, the \textit{Beer2Vec} model performed very well on those beers. Most beers are in line with the style of the base beer. On the other hand, the baseline performs very poorly. In the case of the \textit{Catnip}, only one recommendation out of three was an \textbf{IPA}. The others are either a \textbf{black lager} (\textbf{crisp}) or a \textbf{Gose}  (\textbf{sour}), which is totally off the expected flavor profile. The conclusion is similar for the two other beers, where the PCA model recommended \textbf{Stouts} for the \textbf{Pilsner}, and a \textbf{Lager} and \textbf{Bitter ale} for the \textbf{Stout}. The \textit{Beer2Vec} model recommended, unsurprisingly, beer very close to the requested type. We experimented with the PCA with a higher number of principal components, and the results were as underwhelming as they are presented in Table \ref{tab:baseline_analysis}.

\subsection{Recommendation Quality}

The second element to validate is the value of the recommendation given by our model. 
Even if the \textit{Beer2Vec} recommendation from Table~\ref{tab:baseline_analysis} were adequate, two elements are worth discussing. The first one is the \textbf{wheat beer} recommendation for the \textbf{Pilsner}. \textbf{Wheat beer} has a usual \textbf{ester} character (\textbf{banana}) that is unwanted in a \textbf{Pilsner}. However, the \textit{Joufflue} contains lemon, hinting to the \textbf{crisp} mouthfeel common to every \textbf{Pilsner}. The second one is the \textit{SpécialB}, a \textbf{Belgian Quadrupel} recommended for the \textbf{Stout}. Belgian beers have a highlighted \textbf{fruity} profile that is not usually found in \textbf{Stout}. Upon investigation, all those beer had a strong \textbf{Chocolate}, \textbf{Coffee}, \textbf{Toffee} taste profile.

These two observations support the hypothesis that flavors are not one-hot encoding: most of them overlap, and our models capture this uncertainty and subjectivity in flavor detection by design.

\subsection{Flavors Embedded in Beer Embeddings}

To further support our point, we extracted other recommendations to ensure extensive flavor tag coverage. Table~\ref{tab:more_b2v} presents five additional beers with two recommendations, each accompanied by the recommended beer's prevalent flavors. To obtain the most prevalent tags, we computed the dot product between the beer's vector representation $\mathbf{b}$ and every flavor representation, i.e., $\mathbf{F}$. We kept the three vectors with the highest score.
This table's impressive results confirm the flavor clustering intuition from Figure \ref{fig:2d_taste_vector}. One must note that, even if the prevalent taste profiles are not the same for all recommendations, the various aspects of the base beer's full profile are present in the proposed beer and would, in the aggregate, compose the vector of the base beer style.

\begin{table}
    \centering
    \begin{tabular}{c|c}
    \toprule
         \textbf{Base Beer + Recommendation} & \textbf{Prevalent flavour} \\
        
        \midrule 
        \textbf{Ta Meilleure  (IPA)} & \\
        Les Têtes Boisées (Grisette) & Refreshing, Fruity \\
        Moralité (IPA - American) & Hoppy, Fruity \\
        
        \midrule
        \textbf{IPA Du Nord-Est (NEIPA)} & \\
          Ta Plus Meilleure (NEIPA) & Peachy, Hoppy\\
         L'Escouade B-13 (NEIPA) & Juicy, Hoppy \\
        
        \midrule
        \textbf{Belle Saison (Saison)} & \\
        Benedictus (Belgian Tripel) & Sweet, Floral  \\
        Blanche Beaumont (Wheat beer) & Yeasty, Dry   \\
         \midrule
         \textbf{Rouge de Mékinac (Flanders Red Ale)}  & \\
          Genèse Sauvignon (Belgian Tripel)   & Fruity, Funky \\
          Solstice d'été (Berliner Weisse)  & Sour, Fruity \\
         \midrule
         \textbf{Hudson Bitter (English Bitter)} & \\ 
         Boréale Rousse (Red Ale) & Smooth, Malty \\
         Doppelsticke (Altbier) & Caramel, Bitter \\
        \bottomrule
    \end{tabular}
    \vspace{0.2cm}
    \caption{Most Similar Beers According to \textit{Beer2Vec} with their respective prevalent flavors.}
    \label{tab:more_b2v}
\end{table}

This experiment shows that our method allows us to align vectors so that aggregation (beer) can be explained by its constituents (flavor).

\subsection{Flavour Arithmetics}
One final evaluation scheme would be to verify how flavor arithmetic works within our framework. To achieve this, we start from one beer vector $\mathbf{b}_a$ we subtract the flavor vector $\mathbf{f}_i$ and add the flavor vector $\mathbf{f}_j$ such that $\mathbf{\tilde{b}}_a = \mathbf{b}_a - \mathbf{f}_i + \mathbf{f}_j$. We then find the most similar beer to $\mathbf{\tilde{b}}_a$ in our database. Table \ref{tab:flavour_artimetic} presents 5 examples of flavour arithmetics.

As one can see, the results are not conclusive enough to assert that our algorithmic method works flawlessly. Some results are very interesting, such as the first three examples. The last two are not exactly where they were expected. For instance, The fourth example should be closer to an \textbf{Oud Bruin} or \textbf{Red Flander}, which are  two \textbf{sweet} and \textbf{sour} beers. However, the \textbf{American Wild Ale}, a \textbf{dry} and \textbf{sour} beer, is acceptable. The last one, \textit{Jetlag} does not work at all. The brown ale, a very English type, is far from the added \textbf{belgiany} flavor. One probable cause of the problem is the scarcity of the \textbf{belgiany} tag and the small number of check-in for the \textit{Jetlag} beer.

\begin{table*}[h!!!!!!!!!!!!!!!!!!]
    \centering
    \begin{tabular}{c|c|c|c}
    \toprule
        $\mathbf{b}_a$ & $\mathbf{f}_i$ & $\mathbf{f}_j$ & $\mathbf{\tilde{b}}_a$  \\
        \midrule
        Catnip  (IPA) & Tropical & Caramel & La Marquis De Vaudreuil (Brown Ale - English) \\
        Porter Baltique (Porter) & Coffee & Brett & Wallonade (Belgian Golden)\\
        Full Time I.P.A. (IPA) & Malty & Tropical & Mosaïka (Pale Ale - American)\\
        Sentinelle (Kölsch) & Clean & Sour & Gaspésie Sauvage (American Wild Ale)\\
         Jetlag (Helles) & Crisp & Belgiany & La Marquis De Vaudreuil (Brown Ale - English) \\
         \bottomrule
    \end{tabular}
        \vspace{0.2cm}
    \caption{Illustration of flavors arithmetic. Given a beer vector $\textbf{b}_a$, we subtract one of its flavor $\textbf{f}_i$ and add a new flavor $\textbf{f}_j$ to obtain a new abstract beer vector, $\hat{\textbf{b}}_a$. This vector is then used to retrieve the most similar beer in the beer dataset.}
    \label{tab:flavour_artimetic}
\end{table*}

\subsection{External User Validation}
Our validation scheme relies heavily on the expertise of a brewer. We encourage external users to validate these vectors' quality by trying the \textit{beer2vec} web application for themselves. The reader can find the application at the following address \url{www.anonymous.ca}~\footnote{Website will be publicly available upon acceptance.}.

\begin{figure}[h!!!!!!!]
    \centering
    \includegraphics[width=0.47\textwidth, frame]{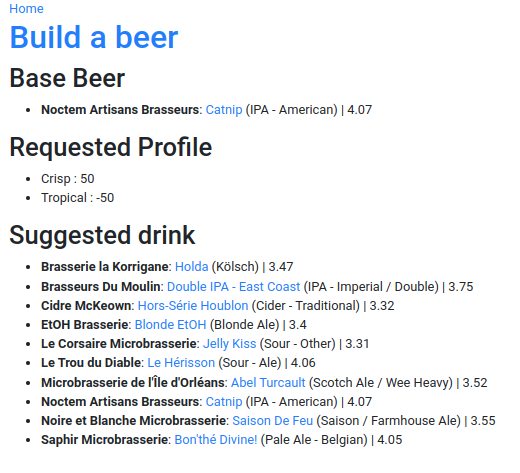}
    \caption{The web interface where the users can build themselves a beer. They begin by selecting a ``base beer'', and then add and/or remove selected flavors accordingly. A suggested list of drinks is then provided by our model, along with its overall appreciation among beer geeks.}
    \label{fig:flavorarithm}
\end{figure}

This application contains the \textit{beer2vec} database used for this paper and the following use cases:

\begin{itemize}
    \item a user can select his favorite beer from a list and find the most similar and dissimilar ones;
    \item A user can select a beer and obtain its associated flavors;
    \item A user can apply flavor vector arithmetic on a base beer (as presented in Figure~\ref{fig:flavorarithm}) and obtain a recommendation.
\end{itemize}

Although the beers contained in the dataset are limited to the available data points we had for our experiment, additional beer can be easily obtained with more beer check-ins and recalculating the flavor vectors using the training procedure discussed in Section~\ref{sec:model}

\section{Conclusion and Future Work}
\label{sec:conclusion}
This paper presents the {\em beer2vec} application that allows users to obtain beer recommendations based on flavor analogies. We presented the algorithm through an analysis of a dataset and presented relevant examples of its functionality.

We intend to expand our application to identify similar breweries, providing an insightful first step to our analogies and helping, for example, the touristic industry to develop personalized beer routes for specific tasting types.

We believe this paper opens the exploitation of formal and standardized tasting cards such as WSET\footnote{\url{https://www.wsetglobal.com/}} to infer flavor vectors. Many official tasting practices, such as sommeliers from governmental alcohol agencies, have this data readability for many products.

\bibliographystyle{splncs04}
\bibliography{main_nlpir}
%

\end{document}